\documentstyle[mprocl,psfig]{article}

\def\Journal#1#2#3#4{{#1} {\bf #2}, #3 (#4)}

\def\PRD{{\em Phys. Rev.} D}

\def\be{\begin{equation}}
\def\ee{\end{equation}}
\def\bea{\begin{eqnarray}}
\def\eea{\end{eqnarray}}

\begin{document}

\title{New Coordinate Systems for Axisymmetric Black Hole Collisions}

\author{ S. R. Brandt$^1$, P. Walker$^{1,2}$, P. Anninos$^3$}

\address{
${}^{(1)}$ Albert--Einstein--Institut,
Schlaatzweg 1, 14473 Potsdam, Germany \\
${}^{(2)}$ Department of Physics,
University of Illinois, Urbana, Illinois 61801 \\
${}^{(3)}$ NCSA,
Beckman Institute, 405 N. Mathews Avenue, Urbana, Illinois, 61801 
}


\maketitle\abstracts{
We describe a numerical grid generating procedure to construct new 
classes of orthogonal coordinate systems that are specially adapted 
to binary black hole spacetimes. The new coordinates offer an
alternative approach to the conventional \v{C}ade\v{z} coordinates,
in addition to providing a potentially more stable and
flexible platform to extend previous calculations of binary
black hole collisions.
}

\section{Introduction}
The two--body problem is a milestone in numerical relativity
that has not yet been attained.
Even the relatively simple case of the head--on
collision of two equal mass, non-rotating black holes remains
uncertain for black holes that are initially separated
by large distances. By exploiting
the 2--dimensional nature of the problem, the obvious
first choice of coordinates in which to solve
the Einstein equations for the axisymmetric collision of two
black holes are the cylindrical ones. However,
these coordinates suffer from a code--crashing axis instability
that can be suppressed (somewhat) by adding a shift vector
to maintain a diagonal metric, but at the expense of introducing steep
shear features in the solutions that cannot be maintained
in a stable manner. An alternative approach,
developed more than 20 years ago,
is to perform the evolutions using curvilinear \v{C}ade\v{z} 
coordinates\cite{cadez}. However, these evolutions are also highly
unstable due to the saddle--point singularity in the coordinate
system. Two methods that have proven to be more successful
invoke elements of both the
cylindrical and \v{C}ade\v{z} coordinates\cite{dcse,paper3}.
These evolutions run well when the black holes are
restricted to separations less than about 
13$M$, where $M$ is the single black hole mass.
However, even the most accurate of these methods\cite{paper3}
is only a partial success when applied to the Misner
initial data since the code
cannot evolve data with large separation parameters.

In this paper, we present an alternative 
approach to the axisymmetric
binary black hole collision by constructing two new classes
of orthogonal body--fitting coordinate systems, specially
adapted to the 2--body problem.

\section{A New Approach}
While saddle--point singularities are unavoidable
in generating the appropriate
body--fitting coordinates for two disconnected domains,
it is reasonable to suppose that problems
associated with these singularities can be minimized
if the saddle--points were moved from the origin (as in the
\v{C}ade\v{z} grid, Fig. 1) to either the
``south pole'' (class I coordinates, Fig. 2), or to the ``north
pole'' (class II coordinates, Fig. 3) of the top hole.  
By setting the lapse to zero on the black hole
throats, the system is prevented from evolving in regions
near the saddle--points and this contributes to the
stability of evolutions in the rest of the spacetime.

\noindent
\begin{tabular}{ccc}
\psfig{figure=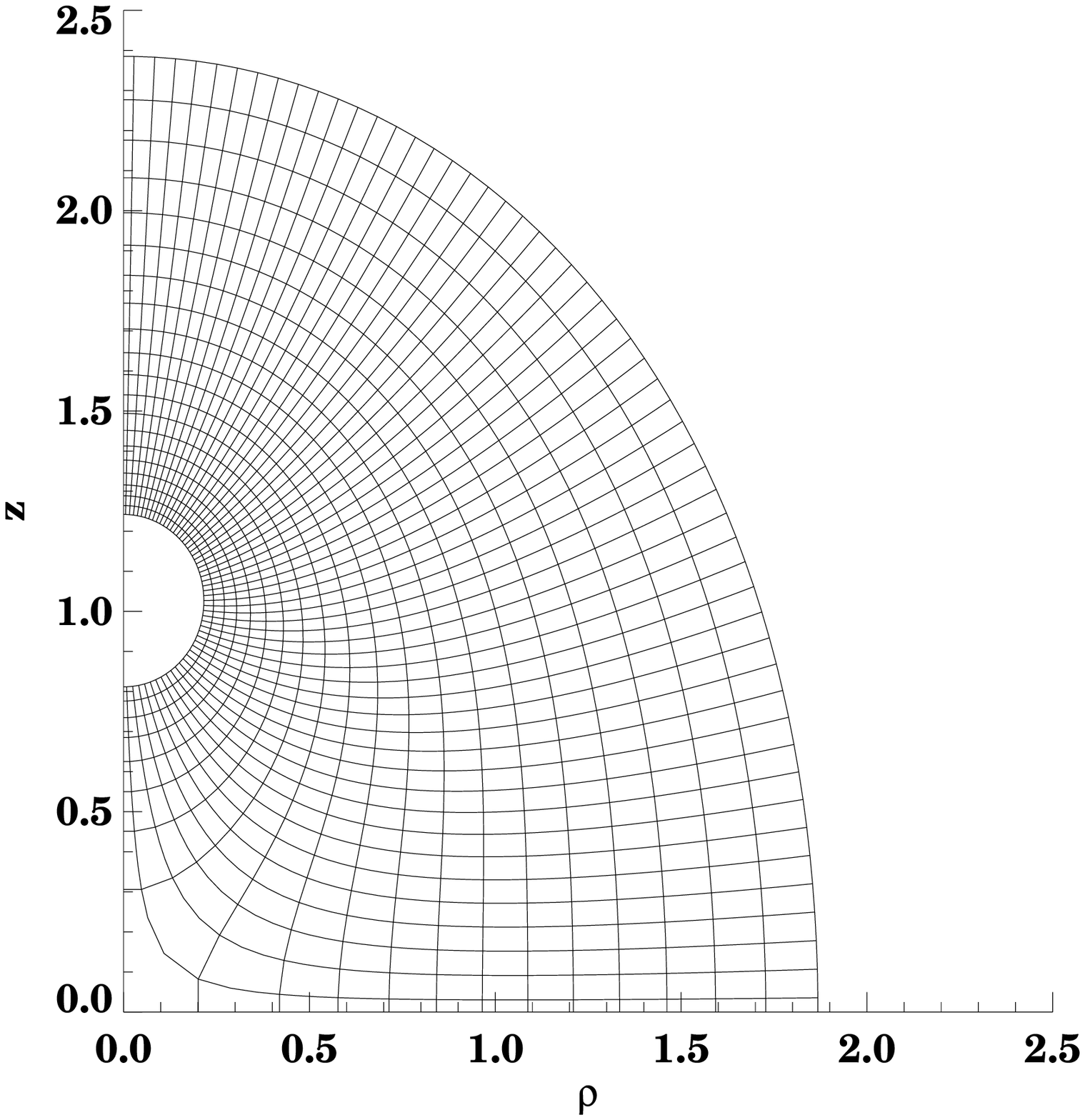,height=4.0cm,width=3.6cm} &
\psfig{figure=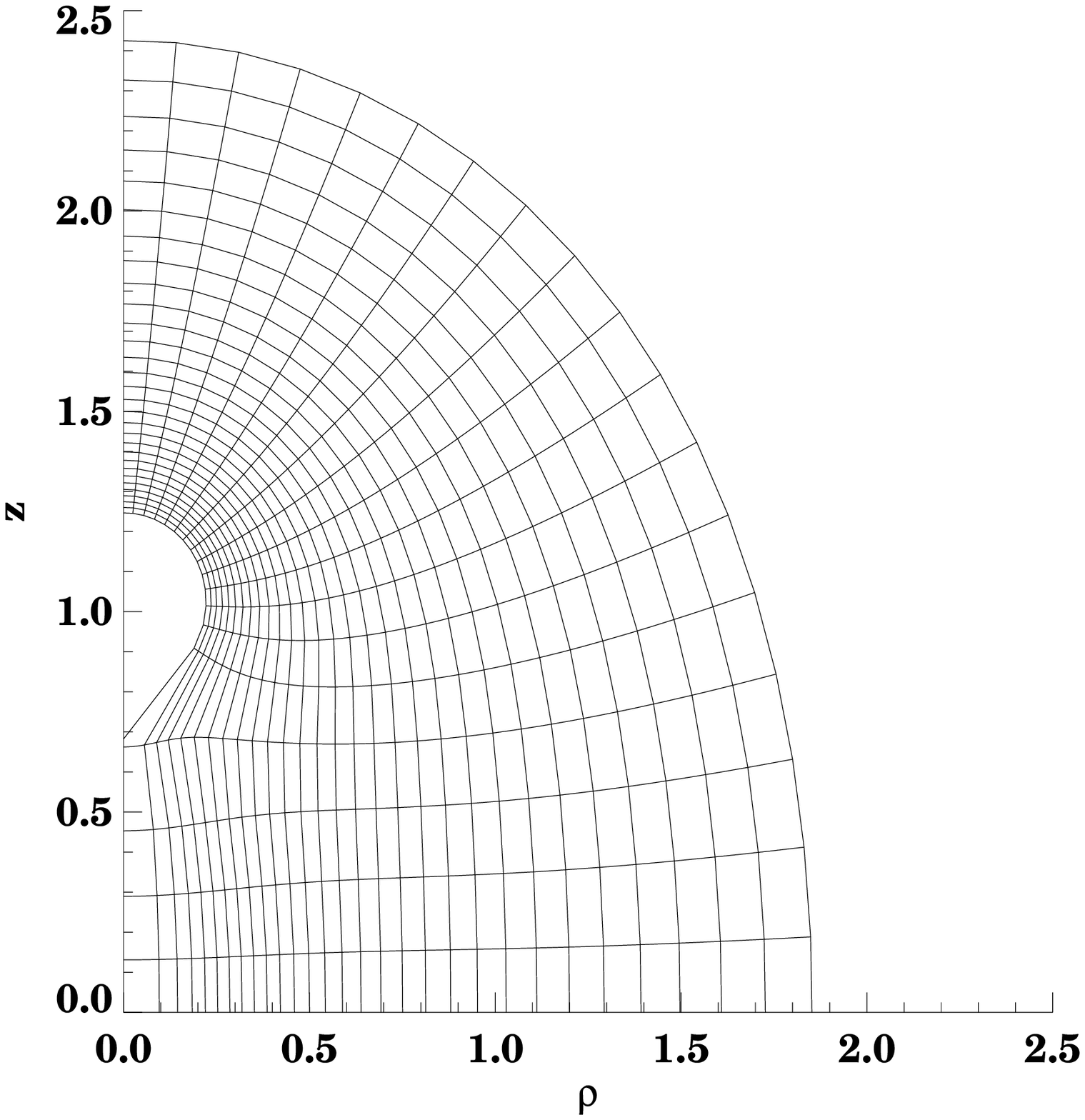,height=4.0cm,width=3.6cm} &
\psfig{figure=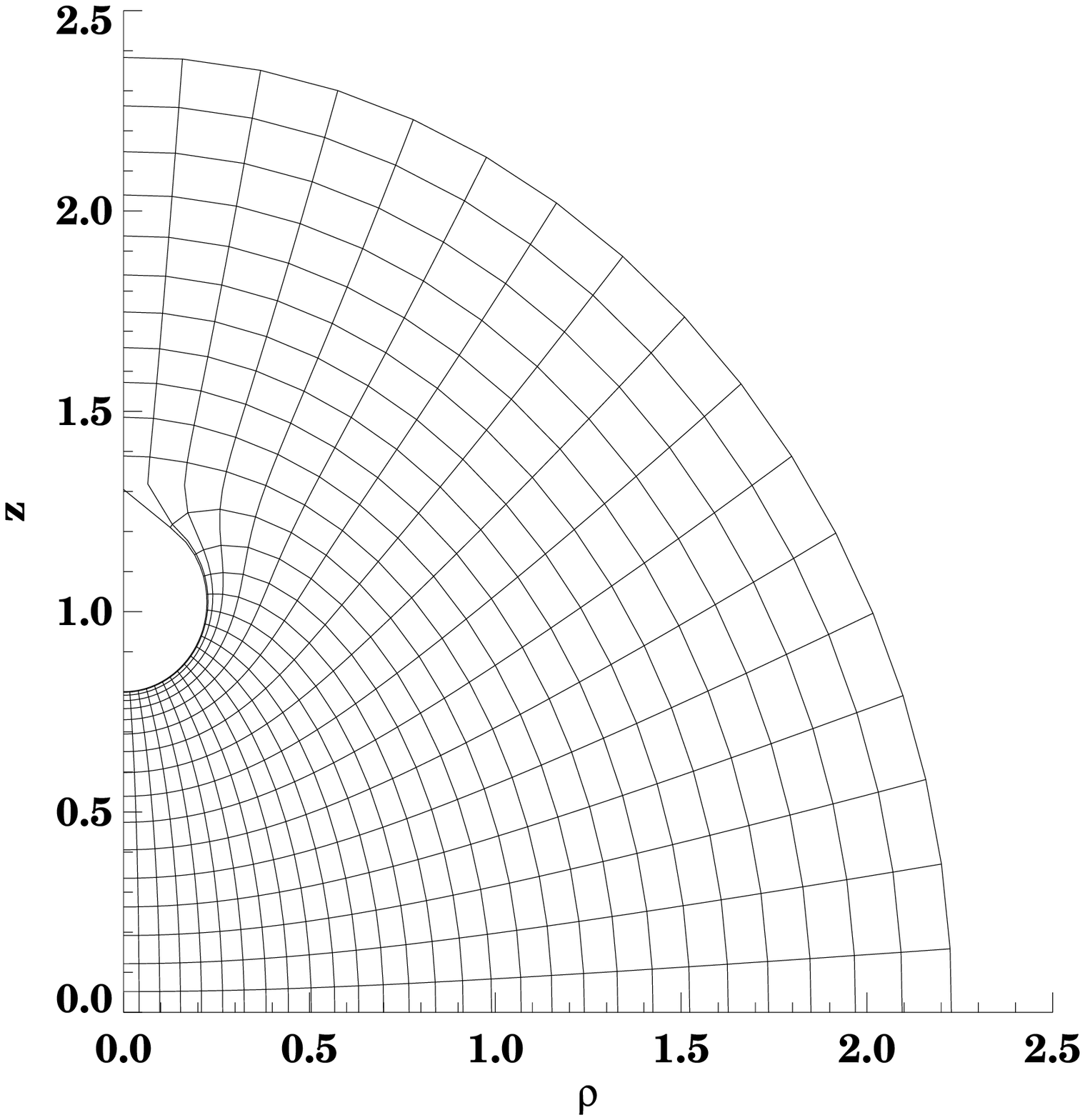,height=4.0cm,width=3.6cm} \\
\begin{minipage}{3.6cm}
\center\footnotesize Fig. 1: \v{C}ade\v{z} grid.
\end{minipage}&
\begin{minipage}{3.6cm}
\center\footnotesize Fig. 2: Class I grid.
\end{minipage}&
\begin{minipage}{3.6cm}
\center\footnotesize Fig. 3: Class II grid.
\end{minipage}
\end{tabular}
\vskip8pt

Our grid generation procedure is based on
a semi--analytic approach: In the class I case,
we construct an analytic {\it radial} coordinate
specially designed to enforce the proper boundary conditions.
The companion angular coordinate is computed numerically by
integrating a set of ODEs that satisfy the orthogonality
conditions for each coordinate line.
In the class II case, we construct
an analytic {\it angular} coordinate and evaluate the
companion radial coordinate numerically.
For example, the class I radial coordinate is
constructed from three separate distance measures
\begin{eqnarray}
r_1 &=& \sqrt{\rho^2+\left(z-z_0\right)^2}-a , \\
r_2 &=& \sqrt{\rho^2+\left(z+z_0\right)^2}-a , \\
r_3 &=& \left(\sqrt{\rho^2+\left(z-z_0+a\right)^2}
+ \sqrt{\rho^2+\left(z+z_0-a\right)^2}- 2 z+2 a\right)/2 ,
\end{eqnarray}
where $a$ is the radius of the black holes,
and $z = \pm z_0$ are the locations of the black hole centers.
$r_1$ and $r_2$ represent distances from the 
two throat surfaces to a point $(\rho,z)$ in conformal space.
The third radius is an ``elliptic distance''
from the central line segment connecting the two holes.
Each of the different radii are
zero on different parts of the inner spectacle--shaped boundary
composed of the two throats and the line segment connecting them.
When combined appropriately, ie.
$r = 3 r_1 r_2 r_3/(r_1 r_2 + r_2 r_3 + r_1 r_3)$,
they form a coordinate that is zero along
the entire inner boundary and becomes spherical at
large distances.

\section{Evolutions}
To date, we have performed several evolutions of black hole
collisions with the new class I coordinates.
Our preliminary results indicate that the extracted waveforms
are less sensitive to the grid and coordinate patch parameters 
(which we continue to use over the first few radial zones)
than in the \v{C}ade\v{z} evolutions.
The evolutions are also able to run for longer times
with large initial separations of the black holes.
Agreement between the different codes is generally good for
moderately separated black holes
($\mu \le 2.5$ in the Misner parameter). 
A comparison of the dominant $\ell = 2$
waveforms between the \v{C}ade\v{z}
and class I grids is presented in Fig. 4
for the $\mu=2.2$ case. We continue
to investigate the further separated black hole cases,
as well as the class II system, and will
report the results in a more comprehensive paper\cite{newpaper}.

With the new class I grids, it is possible to
track the event horizon and locus of null generators\cite{eventh1}
more accurately, since their geometries conform
more closely to the new coordinates than the \v{C}ade\v{z} ones
(which are singular at the critical merger point).
The embedding of the event horizon is shown in Fig. 5
for $\mu = 2.2$.
We note that the spacing between the embedded horizons
is not arbitrary (as it was in the \v{C}ade\v{z} case),
due to the shapes of the horizon and locus in the 
class I coordinate system.

\section{Conclusions}
We have developed two new classes of 
coordinate systems and demonstrated the applicability 
of the class I type in
actual numerical evolutions of colliding black holes.
The new coordinates appear capable to improve on existing
axisymmetric calculations of two black hole collisions.  The
evolutions are more stable, the waveforms less sensitive 
to patch parameters, and the
embedding of the horizon can be determined more 
precisely and with greater ease.
We continue to develop these methods for
class I and II grids, and for evolutions
of black holes with large initial separations.

\vskip10pt
\noindent
\begin{tabular}{ccc}
\psfig{figure=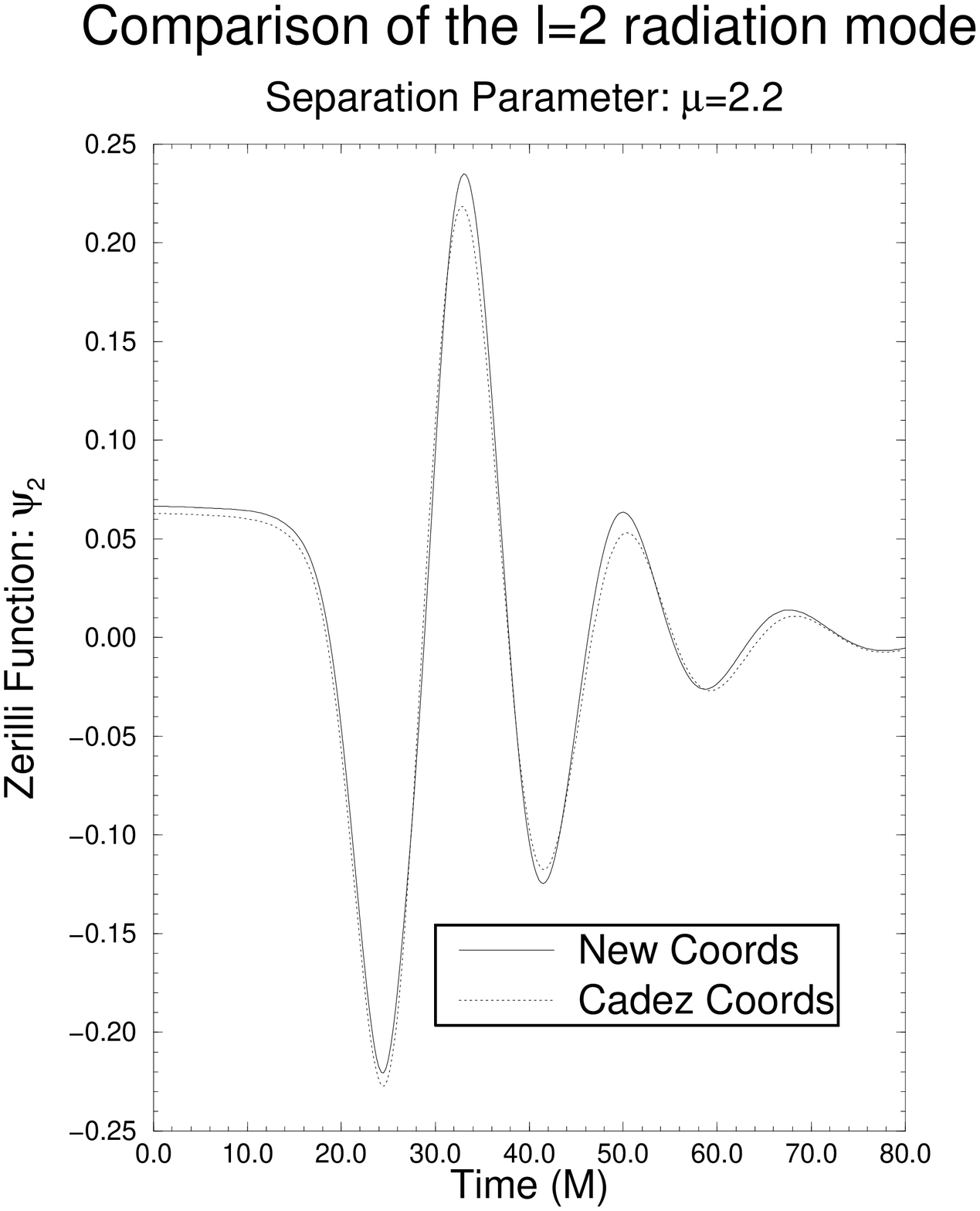,height=5.9cm,width=5.9cm} &
\psfig{figure=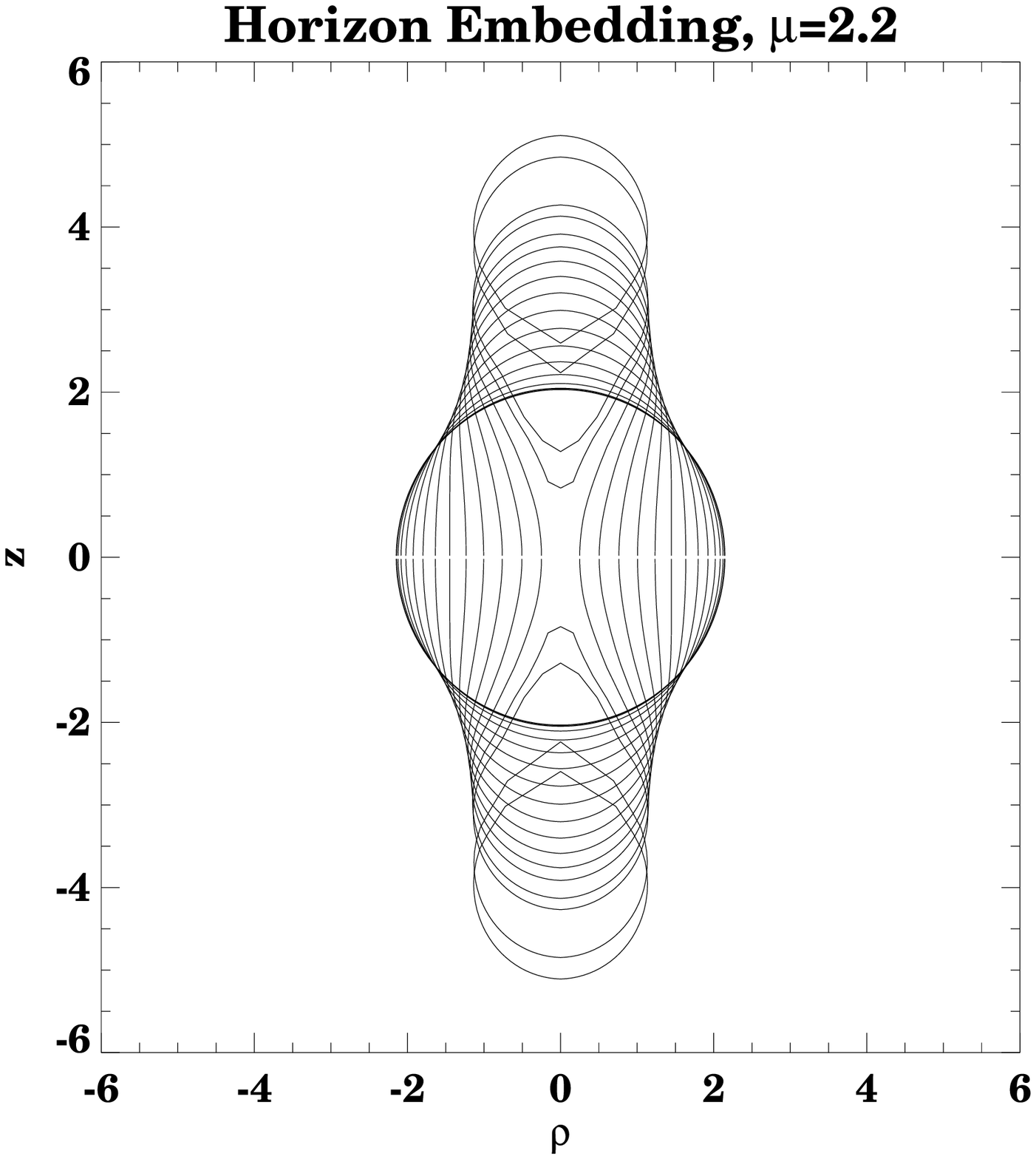,height=5.9cm,width=5.9cm} \\
\begin{minipage}{5.9cm}
\center\footnotesize Fig. 4: 
       $\ell=2$ waveform comparison between the class I 
       and \v{C}ade\v{z} grids.
\end{minipage}&
\begin{minipage}{5.9cm}
\center\footnotesize Fig. 5: 
       Embedding showing the merger of the class I event horizons.
\end{minipage}
\end{tabular}


\section*{References}
\begin{thebibliography}{99}

\bibitem{cadez}
A. Cadez, Ph.D. thesis, University of North Carolina at Chapel Hill,
(1971).

\bibitem{dcse}
L. Smarr, A. Cadez, B. DeWitt and K. Eppley,
\Journal{\PRD}{14}{2443}{1976}.

\bibitem{paper3}
P. Anninos, D. Hobill, E. Seidel, L. Smarr, and W.--M. Suen,
\Journal{\PRD}{52}{2044}{1995}.

\bibitem{newpaper}
P. Anninos, S. Brandt and P. Walker,
in preparation.

\bibitem{eventh1}
J. Libson, J. Mass\'o, E. Seidel, W.-M. Suen, and P. Walker, 
\Journal{\PRD}{53}{4335}{1996}.


\end{thebibliography}
\end{document}